\documentclass{emulateapj}

\slugcomment{Accepted in AJ}

\begin{document}

\title{Low-Amplitude Variables: Distinguishing RR Lyrae stars from 
       Eclipsing Binaries.}

\author{T.D.Kinman\altaffilmark{1}}           
\affil{NOAO, P.O.Box 26732, Tucson, Arizona 85726, USA}

\author{ Warren  R. Brown}                             
\affil{Smithsonian Astrophysical Observatory, 60 Garden St.,    
 Cambridge, MA 02138, USA}

\altaffiltext{1}{ 
 The NOAO are operated by AURA, Inc.\ under cooperative 
agreement with the National Science
Foundation.}

\begin{abstract}

  It is not easy  to identify and classify low-amplitude variables, but it
  is important that the classification is done correctly. 
  We use photometry and spectroscopy to classify low-amplitude variables 
  in a 246 deg$^{2}$ part of the Akerlof et al. (2002) field. Akerlof and
  collaborators found that 38\% of the RR Lyrae stars in their 2000 deg$^{2}$
  test field were RR1 (type $c$). This suggests that these RR Lyrae stars 
  belong to an Oosterhoff Type II population while their period distribution
  is primarily Oosterhoff Type I. Our observations support their RR0 (type $ab$)
  classifications, however 6 of the 7 stars that they classified as RR1 
  (type $c$) are eclipsing binaries. Our classifications are supported by
  spectroscopic metallicities, line-broadening and Galactic rotation 
  measurements. Our 246 deg$^{2}$ field contains 16 RR Lyrae stars that are
  brighter than m$_{R}$ = 14.5; only four of these are RR1 (type $c$). This
  corresponds to an Oosterhoff Type I population in agreement with the period
  distribution.

\end{abstract}


\section{Introduction}

 Low-amplitude variables are difficult to identify and classify, yet determining
 their true nature is astrophysically important. 
 Lee and Carney (1999) and Miceli et al. (2008) have shown that the Oosterhoff
 Type I (Oo I) and the Oosterhoff Type II (Oo II) {\it field} RR Lyrae stars
 have different galactic distributions and hence, presumably, different 
 origins. 
 Oosterhoff Types are defined not only by the period distributions of their 
 RR0 (type {\it ab}) and RR1 (type {\it c}) variables but also by the relative 
 numbers of these two RR Lyrae types.
 Thus the RR1 (type {\it c}) comprise 45\% of all the RR Lyrae stars in 
 the Oo II globular clusters but only 20\% in the OoI globular clusters 
 (Clement et al.  2001). In recent large surveys for field RR Lyrae stars,
 RR1 variables comprised 17\% of the total in the QUEST Survey (Vivas et al. 
 2004) and 22\% of the total in the SDSS Stripe 82 Survey (Watkins et al. 
 2009) indicating Oo I populations in both cases. 

 Akerlof et al.. (2000) (hereafter AAB), on the other hand, study a 2000 deg$^{2}$
 field of the $ROTSE$ All-Sky Survey 
 \footnote{Hereafter called the $ROTSE1$ survey. We refer
 to the complete Northern Sky Variability Survey (Wozniak et al.,  2004) as
 the $NSVS$.}, and find that 38\% 
  of their RR Lyrae stars are RR1 (type {\it c}). Problematically, this 
  suggests an Oo II population although the period distribution
 shown in AAB (Fig. 8) is that of a  predominantly Oo I population. 
 The $QUEST$, $SDSS$ and $ROTSE1$ surveys all use CCD photometry 
 which (unlike the older photographic surveys) can be used to detect the
 relatively low-amplitude RR1 (type $c$) variables with reasonable completeness. 
  The $ROTSE1$ survey, for example, lists variables with amplitudes
 as low as 0.1 mag---significantly lower than the amplitudes of most
 RR1 (type {\it c}). Amrose \& McKay (2001) did not include the RR1 
 (type {\it c}) in their discussion of the {\it ROTSE1} RR Lyrae variables 
 because they considered their classification less robust than that of the 
  RR0 (type {\it ab}). 
 We also note that the stars that AAB classify as RR1 (type $c$) tend to be 
 brighter and to have lower reduced proper motions than those that they classify
 as RR0 (type $ab$); this suggests that we should review the classification
 of the lower amplitude RR Lyrae stars in the $ROTSE1$ survey.

 Recently, Hoffman et al. (2009) have re-classified the brighter variables
 in the $NSVS$ using Fourier coefficients. Even in this bright sample, they
 say `` without color information and higher precision photometry, the W UMa
 and RRc variables cannot easily be differentiated" and in their summary
 they cite the W UMa/RRc degeneracy problem as a major source of 
 misclassification. Blomme et al. (2010) mention the problem in their 
 discussion of the new Kepler data and it surely will be a problem in 
 the pipeline analysis of Pan-STARRS and LSST variable star data. 

  In this paper we consider a 246 deg$^{2}$ field near the North Galactic
 Pole (NGP) defined by 186.$^{\circ}$5$<$ R.A.$<$ 204.$^{\circ}$0 and 
 +23.$^{\circ}$0$<$ Dec.$<$+39.$^{\circ}$0 (Fig. 1). This is part of the
 2000 deg$^{2}$ test field $ROTSE1$. In addition to discussing the data given by
  AAB, we use the General Catalogue of Variable Stars ($GCVS$) (Samus  2005) 
 and various sources discussed in the Appendix (A). 

     AAB identified 33 RR Lyrae stars in this field; one of these
(J130441.22+381804.50) is the same as J130441.18+381805.1 and is ignored.
 AAB classified 9  as RR1 (type {\it c}) and 
 23 as RR0 (type {\it ab}). Two of these RR1 (U Com and VW CVn) are given in
 the $GCVS$. U Com has been studied in detail (Bono et al.  2000) and its
 classification is secure.  VW CVn, however, is classified as 
 an eclipsing binary in both the $GCVS$ and SIMBAD. It is also included in a
 recent catalogue of eclipsing variables (Malkov et al., 2006). 
 AAB classify VW CVn as RR1 (type {\it c}) and it is clear from the light
 curve given by Agerer and Berthold (1994) that this classification is 
 the correct one  with a period of 0.425 days and not (as originally 
 thought) an eclipsing system with  twice this period. {\it This confusion
 over the nature of a 12th magnitude star shows that the classification of
 RR1 variables from low S/N data is not trivial}. 

 The 21 $ROTSE1$ RR 
 Lyrae stars in our field that are in the $GCVS$ are listed in Table~1. The 
 $ROTSE1$ data for the other 11 RR Lyrae variables are given in Table~2.
 Further information on these stars is given in the notes to this table.
 Apart from VW CVn, we assume that the $GCVS$ classifications given in 
 Table 1 are correct but that the classifications of the stars in Table~2 
 need to checked  through  new photometry, spectroscopy and a 
 rediscussion of the $NSVS$ photometry for these variables.

\begin{deluxetable*}{ ccccccccc }
\tablewidth{0cm}
\tabletypesize{\scriptsize}
\setcounter{table}{1}
\tablecaption{ $ROTSE1$ RR Lyrae variables that are identified in the $GCVS$.}
 
\tablehead{ 
\colhead{  $ROTSE1$ ID  } &
\colhead{  type \tablenotemark{a} } &
\colhead{ Period                } &
\colhead{ m$_{R}$ \tablenotemark{b} } &
\colhead{ A               \tablenotemark{c}  } &
\colhead{                                    } &
\colhead{ ID   \tablenotemark{d}  } &
\colhead{ type    \tablenotemark{e}  } &
\colhead{ Period       \tablenotemark{f}  } \\
                  &   &(days) &    &(mag.)&  &      &    &  (days)       \\
}
\startdata
                   &    &     &     &      & &      &    &                \\
J122907.47+343850.0&rrab&0.55868&13.17&1.0& &RR CVn&RRab&0.55860        \\
J123245.60+270145.7&rrab&0.58668&12.20&1.0& & S Com&RRab&0.58659        \\
J123556.02+371224.9&rrab&0.66847&12.98&0,7& &SV CVn&RRab&0.66806       \\
J123757.13+295805.8&rrab&0.47266&15.02&1.3& &FV Com&RRab&0.47246        \\
J124004.01+273014.0& rrc&0.29278&11.57&0.3& & U Com&RRc &0.29273        \\
J124055.05+370507.0&rrab&0.44169&13.20&0.8& &SW CVn&RRab&0.44165        \\
J124354.34+280114.0&rrab&0.54080&15.11&1.0& &DV Com&RRab&0.54084        \\
J124716.30+351206.0&rrab&0.61849&15.10&1.2& &DS CVn&RRab&0.61843       \\
J125000.65+310824.4&rrab&0.53654&14.75&1.0& &TX Com&RRab&0.53647        \\
J125110.42+325808.3&rrab&0.57463&13.38&0.4& &AP CVn&RRab&0.57465        \\
J125421.57+321433.3&rrab&0.51345&14.11&0.9& &TY CVn&RRab&0.51344        \\
J125455.50+231526.3&rrab&0.54176&14.82&0.9& &BD Com&RR  &0.54161        \\
J125952.34+301432.6&rrab&0.53235&14.75&0.9& &UW Com&RRab&0.53233        \\
J130129.21+320512.3&rrab&0.55198&14.63&1.1& &TZ CVn&RRab&0.55187        \\
J130213.65+241419.6&rrab&0.66163&13.95&0.8& &BF Com&RRab&0.66141        \\
J130507.95+231642.8&rrab&0.46901&13.09&0.9& &RY Com&RRab&0.46895        \\
J131226.95+302117.9&rrab&0.73723&13.64&0.5& &UZ Com&RRab&0.73694        \\
J131703.38+360656.3&rrab&0.67783&15.10&1.0& &DZ CVn&RRab&0.67732     \\
J132942.14+285248.2& rrc&0.42518&12.02&0.4& &VW CVn&EW  &0.85001        \\
J133430.88+291815.5&rrab&0.52353&15.38&1.2& &WW CVn&RRab&0.52340         \\
J133455.38+262700.2&rrab&0.57352&15.30&1.3& &BT Com&RRab&0.57355         \\
                   &  &       &     &   & &      &    &                \\
\enddata

\tablenotetext{a}{Type given by Akerlof et al. (2000) rrab = RR0; rrc = RR1     
 .} 
\tablenotetext{b}{Unfiltered CCD mean magnitude of variable.} 
\tablenotetext{c}{Unfiltered CCD amplitude of variable.   } 
\tablenotetext{d}{Identification in $GCVS$                   } 
\tablenotetext{e}{Variable type given in $GCVS$              } 
\tablenotetext{f}{Period given in $GCVS$                             } 

\end{deluxetable*}

\begin{figure}
\plotone{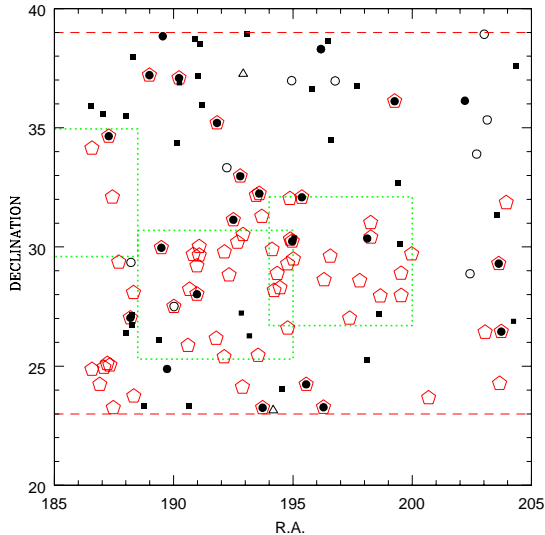}
\caption{ Our field at the North Galactic Pole that contains $ROTSE1$ variables
     lies between Declinations +23.$^{\circ}$0 and +39.$^{\circ}$0
         (shown by the horizontal red dashed lines).
   The RR Lyrae stars listed in the $GCVS$ are shown by large open red
   pentagons. The stars  that were
   discovered by AAB are shown (using their classifications)  by the
   following black symbols: filled circles RR0 (type $ab$); open circles
   RR1 (type $c$); open triangles Delta Scuti stars. The small filled
   squares show the stars that AAB classified as eclipsing binaries.
   The survey fields of Kinman et al. (1966) are outlined by green dotted
   lines.
\label{Fig1}}
\end{figure}

\begin{deluxetable*}{ ccccccccc }
\tablewidth{0cm}
\tabletypesize{\scriptsize}
\tablecaption{ $ROTSE1$ RR Lyrae variables that are not identified in the $GCVS$.}

\tablehead{
\colhead{  $ROTSE1$ ID  } &
\colhead{  type \tablenotemark{a} } &
\colhead{ Period \tablenotemark{b}                } &
\colhead{ JD Max \tablenotemark{c} } &
\colhead{ D               \tablenotemark{d}  } &
\colhead{ m$_{R}$ \tablenotemark{e}          } &
\colhead{  A   \tablenotemark{f}  } &
\colhead{ Q   \tablenotemark{g}  } &
\colhead{ Note   } \\
                   &    & (days)            &          &   &     &(mag.)&  &           \\
}

\startdata
                   &    &                   &          &   &     &   &  &           \\
J123250.33+292123.6&rrc &0.26217$\pm$0.00003&51244.6032&0.5&11.68&0.1& 6& (1)\\
J123811.00+385028.0&rrab&0.53331$\pm$0.00006&51246.6095&0.2&13.94&1.1& 5& (2)\\
J123854.21+245307.9&rrab&0.62841$\pm$0.00014&51244.9048&0.4&13.24&0.3& 9& (3)\\
J124855.82+331934.5&rrc &0.20415$\pm$0.00002&51244.8268&0.5&12.51&0.2& 5& (4)\\
J125947.50+365843.6&rrc &0.30817$\pm$0.00002&51244.5879&0.2&10.62&0.3& 5& (5)\\
J130441.18+381805.1&rrab&0.68636$\pm$0.00011&51246.8109&0.3&14.03&0.9& 9& (6)\\
J130705.50+365757.1&rrc &0.22348$\pm$0.00002&51244.5912&0.4&11.59&0.2& 5&    \\
J132849.66+360757.1&rrab&0.58633$\pm$0.00016&51246.6046&0.5&14.87&0.8& 6& (7)\\
J133048.36+335353.8&rrc &0.35292$\pm$0.00005&51246.6543&0.5&14.30&0.7& 5& (8)\\
J133204.15+385533.7&rrc &0.28913$\pm$0.00007&51244.8027&0.9&11.16&0.1& 5& (9)\\
J133234.47+351949.5&rrc &0.35983$\pm$0.00006&51244.4898&0.5&13.14&0.3& 6&    \\
                   &  &       &     &   & &      &    &                \\
\enddata

\tablenotetext{a}{Type given by Akerlof et al. (2000) rrab = RR0; rrc = RR1
 .}
\tablenotetext{b}{Period in days (and error) }
\tablenotetext{c}{Julian Date of maximum light (+2400000.) }
\tablenotetext{d}{Accumulated phase error after 3 years. }
\tablenotetext{e}{Unfiltered CCD mean magnitude of variable.}
\tablenotetext{f}{Unfiltered CCD amplitude of variable.   }
\tablenotetext{g}{Quality on scale 10 (excellent) downwards   }

\tablecomments{
 {\bf 01} TYC 1991-1673-1 (Hog et al.  2000);  BD +30~2289                 \\
 {\bf 02} BPS BS 17141-0020 (Beers et al.  1996)                           \\
 {\bf 03} Period = 0.$^{d}$62839, [Fe/H] = --1.2 (Kinemuchi et al.  2006)         \\
 {\bf 04} No. 213 (Slettebak \& Stock  1959); A-F~140 (Sanduleak  1988      \\
 {\bf 05} BD +37~2346; TYC 2534-1296-1 (Hog et al.  2000); A-F~813
         (MacConnell et al.  1993). H\"{a}ggkvist \& Oja (1973) give $V$ =
 10.19, $B-V$ = +0.21 \& $U-B$ = 0.10 appropriate for Main Sequence star. \\
 {\bf 06} BPS BS 16938-0026 (Beers et al.  1996)
  Period = 0.$^{d}$68511, [Fe/H] = --2.2 (Kinemuchi et al.  2006)         \\
 {\bf 07} A-F~930 (MacConnell et al.  1993); BPS BS 16924-0007 (Beers et al.
 1996). Period = 0.$^{d}$58620 (Kinemuchi et al.  2006)                  \\
 {\bf 08} A-F~938 (MacConnell et al.  1993); BPS BS 16924-0010 (Beers et al.
 1996)   \\
 {\bf 09} TYC 3025-723-1 (Hog et al.  2000); BPS BS 17450-0019 (Beers et al.
 1996)  \\
    }

\end{deluxetable*}

\begin{deluxetable}{ ccccc }
\tablewidth{0cm}
\tabletypesize{\footnotesize}
\tablecaption{ Photoelectric photometry of $ROTSE1$ variables in Table 2.}
 
\tablehead{ 
\colhead{  $ROTSE1$ ID  } &
\colhead{ JDH \tablenotemark{a}                } &
\colhead{ $V$ \tablenotemark{b} } &
\colhead{ $(B-V)$         \tablenotemark{c}  } &
\colhead{ Phase \tablenotemark{d}          } \\
}

\startdata
                   &                    &     &                &            \\
J123250.33+292123.6&52407.8184   & 11.422 & 0.378   & 0.691    \\
J123250.33+292123.6&52408.8117   & 11.371 & 0.374   & 0.585    \\
J123250.33+292123.6&52409.6892   & 11.450 & 0.376   & 0.259    \\
J123250.33+292123.6&52411.6874   & 11.414 & 0.380   & 0.071    \\
J123250.33+292123.6&52411.7903   & 11.440 & 0.377   & 0.267    \\
J123250.33+292123.6&52416.7395   & 11.457 & 0.375   & 0.707    \\
J123250.33+292123.6&52417.7894   & 11.419 & 0.367   & 0.709    \\
J123250.33+292123.6&52418.7218   & 11.440 & 0.373   & 0.488    \\
                   &                    &     &            &            \\
\enddata

\tablenotetext{a}{Heliocentric Julian Date  (+2400000.) } 
\tablenotetext{b}{Johnson $V$ magnitude  } 
\tablenotetext{c}{Johnson $(B-V)$ color } 
\tablenotetext{d}{Phase for ephemeris given in text.}

\end{deluxetable}

\begin{deluxetable}{ ccccc }
\tablewidth{0cm}
\tabletypesize{\footnotesize}
\tablecaption{ Tenagra CCD photometry of $ROTSE1$ variables in Table 2.}
 
\tablehead{ 
\colhead{  ROTSE1 ID  } &
\colhead{ JDH \tablenotemark{a}                } &
\colhead{ $V$ \tablenotemark{b} } &
\colhead{ $B$ \tablenotemark{c}  } &
\colhead{ Phase \tablenotemark{d}          } \\
}

\startdata
                   &                    &     &                &            \\
J123250.33+292123.6&54543.8937   &$\cdots$& 11.77  &  0.993    \\
J123250.33+292123.6&54574.7290   &$\cdots$& 11.82  &  0.808    \\
J123250.33+292123.6&54575.6632   &$\cdots$& 11.85  &  0.590    \\
J123250.33+292123.6&54578.7472   &$\cdots$& 11.66  &  0.472    \\
J123250.33+292123.6&54579.6928   &$\cdots$& 11.77  &  0.276    \\
J123250.33+292123.6&54580.7074   &$\cdots$& 11.85  &  0.211    \\
J123250.33+292123.6&54845.0301   &$\cdots$& 11.75  &  0.374    \\
J123250.33+292123.6&54846.0272   &$\cdots$& 11.83  &  0.276    \\
J123250.33+292123.6&54847.0444   &$\cdots$& 11.83  &  0.216    \\
J123250.33+292123.6&54847.0548   &$\cdots$& 11.85  &  0.236    \\
                   &                    &     &                &            \\
\enddata

\tablenotetext{a}{Heliocentric Julian Date  (+2400000.) } 
\tablenotetext{b}{Johnson $V$ magnitude  } 
\tablenotetext{c}{Johnson $B$ magnitude  } 
\tablenotetext{d}{Phase for ephemeris given in text.  }

\end{deluxetable}

\section{Photometry of Variables}

We first observed the variables in Table 2  on 9 nights of a 12-night run
with the 42-inch John S. Hall telescope of the Lowell Observatory using the
Kron aperture photometer and a thermoelectrically 
cooled EMI 6256 photomultiplier.
Landolt standards  (Landolt, 1992) were observed nightly so that the $V$ and
$(B-V)$ are on the Johnson system. These observations (made in May 2002; 
 JD 2452407 $-$ 2452418) are given in Table 3.  Only about ten observations 
 were obtained for each object. Also, although the sky was apparently clear for
 these observations, the scatter in the $V$ magnitudes suggests that it may not
 have always been perfectly photometric. The effect on the $(B-V)$ colors 
  was probably minor in comparison. 
 We therefore obtained further photometric observations
  between 2004 and 2009 using the commercial robotic 
 f/7 0.8-m Ritchey-Chretien telescope of the Tenagra Observatory in Arizona
 (Schwartz  2007). The detector 
 on this telescope was a 1024 $\times$ 1024 SITe CCD. Only the central 
 7-arcmin diameter field (which has excellent cosmetic quality) was used
 and reduced with standard $IRAF$ routines (Tody  1993). On photometric nights
 the data were calibrated using local standards from an earlier 
 program at the North Galactic Pole (Kinman et al., 1994). The variable 
 J123811.00+385028.0 and its comparison star (A) at J123806.03+385049  
  have also been 
 observed by Schmidt (private communication). He found $V$ = 13.848 and 
 $V-R$ = 0.372 for the comparison star while we get $V$ = 13.841  and $B-V$ = 
 0.27 in satisfactory agreement. On non-photometric nights, the magnitudes of 
 the variables were obtained differentially with respect to nearby stars. 
 The positions of these comparison stars and their adopted magnitudes 
 are given in Table 2 in the Appendix (C).  In
 some cases only relatively faint comparison stars were available in the
 7 arcmin field and this limited the attainable accuracy with the relatively
 short exposures that were used. 

\begin{figure}
\plotone{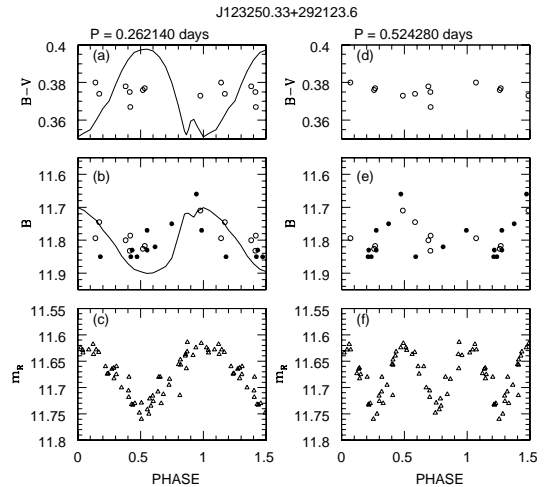}
\caption{ Light curves for J123250.33+292123.6. Open circles show the
     Lowell photoelectric data (2002). Filled circles show the Tenagra 
     CCD photometry (2008, 2009). The open triangles show the light curves
   derived from $NSVS$ photometry (1999, 2000). The plots on the left are
  for a period of 0.262140 days and those on the right are for a period
          of 0.524280 days which is the preferred interpretation. The 
  curves in panels (a) and (b) are those that would be expected if
  the star were an RRc variable; it is seen that they do not fit the data.
\label{Fig2}}
\end{figure}

\begin{figure}
\plotone{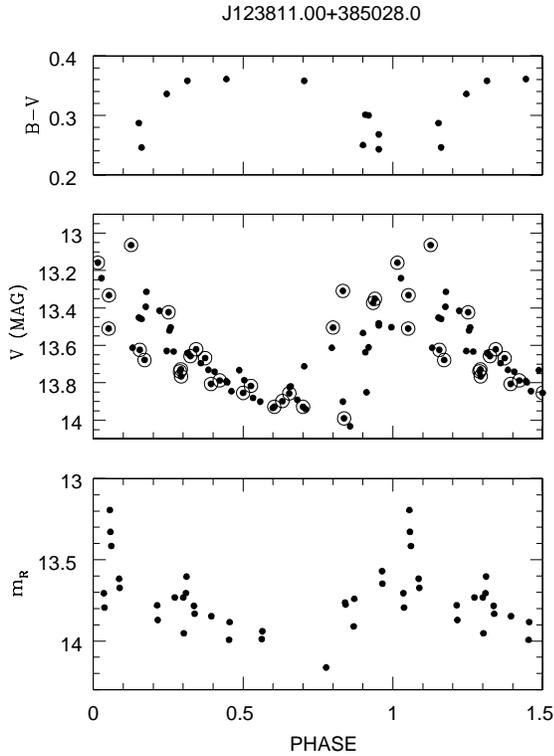}
\caption{ Light curves for  the RR Lyrae star J123811.00+385028.0. 
     Upper panel: $(B-V)$ from Lowell photooelctric data (2002).
     Middle panel: V magnitudes. 
     The Lowell photoelectric data (2002) and the Tenagra CCD data 
      (2004, 2008 \& 2009) are shown by filled circles.
      The CCD data (2009) made available by E. Schmidt are  shown 
          encircled. 
     Lower panel: $m_{R}$ magnitudes from $NSVS$ (1999). 
\label{Fig3}}
\end{figure}

  The photoelectric data of 2002 are given in Table 3 and the Tenagra 
  photometric data are given in Table~4\footnote{Both tables are available
  in their entirety in machine-readable and Virtual Observatory (VO)
  forms in the on-line journal. A portions of each are shown here for guidance 
  regarding their forms and content.}. We also used the Northern Sky 
  Variability Survey ($NSVS$, Wozniak et al.  2004). 
   Typically, this data contains many pairs of
  observations that are separated by about 0.0010 days and which can be
  combined to improve the accuracy with negligible loss of time resolution. 
  We also rejected data if the quoted errors were relatively large.

  We illustrate the problem of using photometry for classification by 
  considering the case of one of the lowest amplitude variables
  (J123250.33+292123.6) (Fig. 2).  Our Johnson $B$ 
  magnitudes are not accurate enough to give an adequate light curve for a
  variable with such a low amplitude (Fig. 2 (b)(e)). The $NSVS$ data
  in $m_{R}$
 \footnote{We use $m_{R}$ for the unfiltered CCD magnitudes of the $NSVS$ 
  survey.}, however, gives a  much better light curve. We compared  12 bright
  RR Lyrae stars that have both $m_{R}$ and Johnson $BV$ light curves and
  found that their $m_{R}$,~$V$ and $B$ amplitudes are in the ratios of  
  1.00, 1.27 and 1.60 respectively. The $m_{R}$--amplitude of 
  J123250.33+292123.6 is roughly 0.125 mag., so if it were  a pulsating star
  we would expect amplitudes of 
  0.20 and 0.041 mag. in $B$ and $(B-V)$ respectively. In Fig.2 (a) and (b)
  we have taken the $(B-V)$ and $B$ curves of the RRc U Com (Heiser,
   1996),  and scaled them to those of J123250.33+292123.6 using the ratio of
  their $m_{R}$ amplitudes; these curves give a poor fit to the data.
   For an eclipsing system, we would expect the 
  $m_{R}$ and $B$ amplitudes to be equal and the color to be constant. 
  Our $B$ photometry is not accurate enough to give a reliable amplitude 
  for this star but the
   small scatter in $(B-V)$ (Fig. 2 (a,d) and the shape of the $m_{R}$
  light curve (Fig. 2 (c,f)),  suggests that J123250.33+292123.6 is
  an eclipsing system with P = 0.52428 days rather than a pulsating star
  with half this period. A summary of its photometric properties is given in Sec. 2.2.

\begin{figure}
\plotone{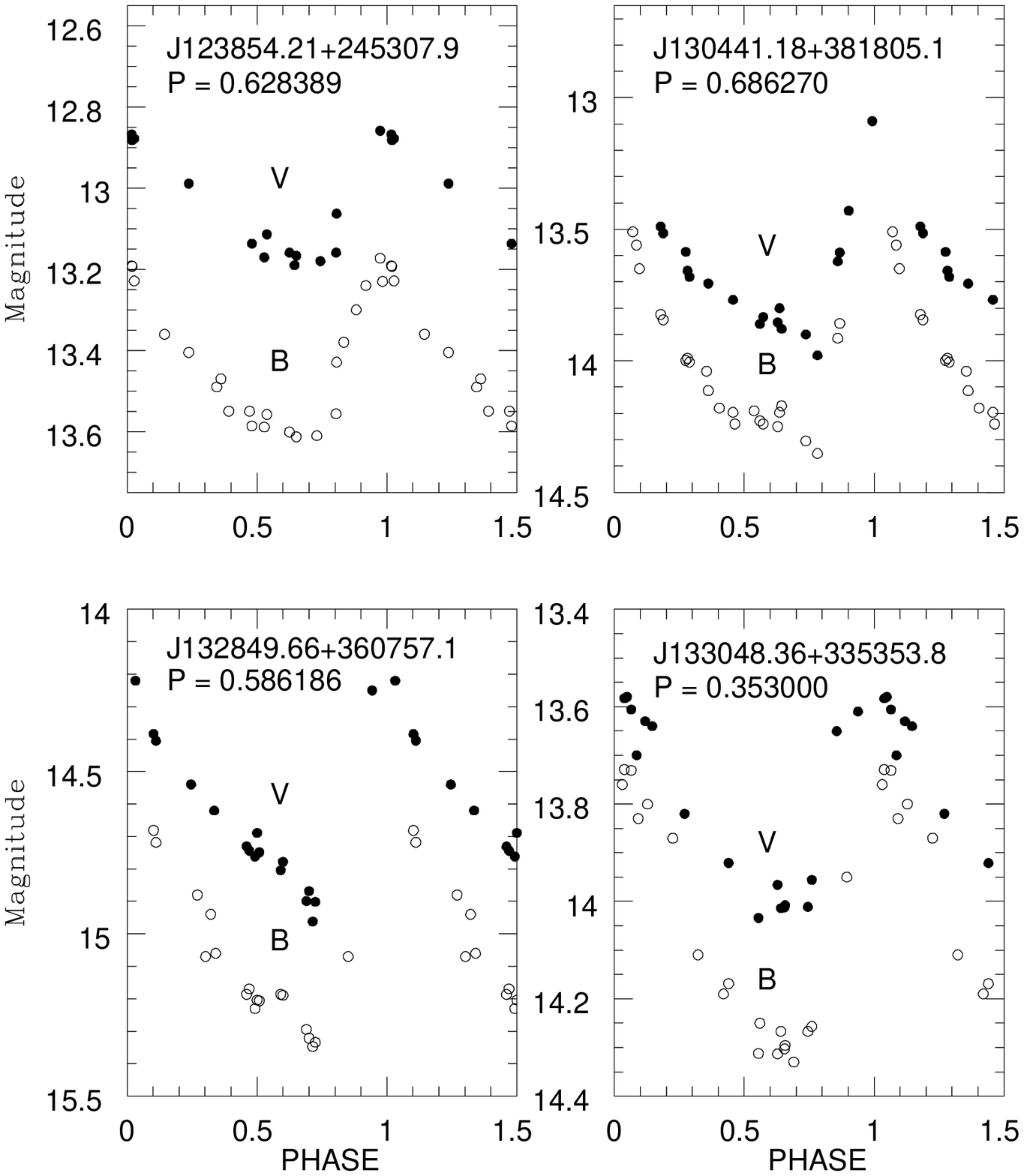}
\caption{ Light curves for J123854.21+245307.9, J130441.18+381805.1, 
         J132849.66+360757.1 and J133048.36+335353.8. They are based on
 both the Lowell photoelectric and Tenagra CCD data and cover the years
    2002, 2004, 2008 and 2009. $V$ magnitudes are shown as filled circles
         and $B$ magnitudes as open circles. All these stars are classified
         as RR Lyrae stars.
\label{Fig4}}
\end{figure}

\begin{figure}
\plotone{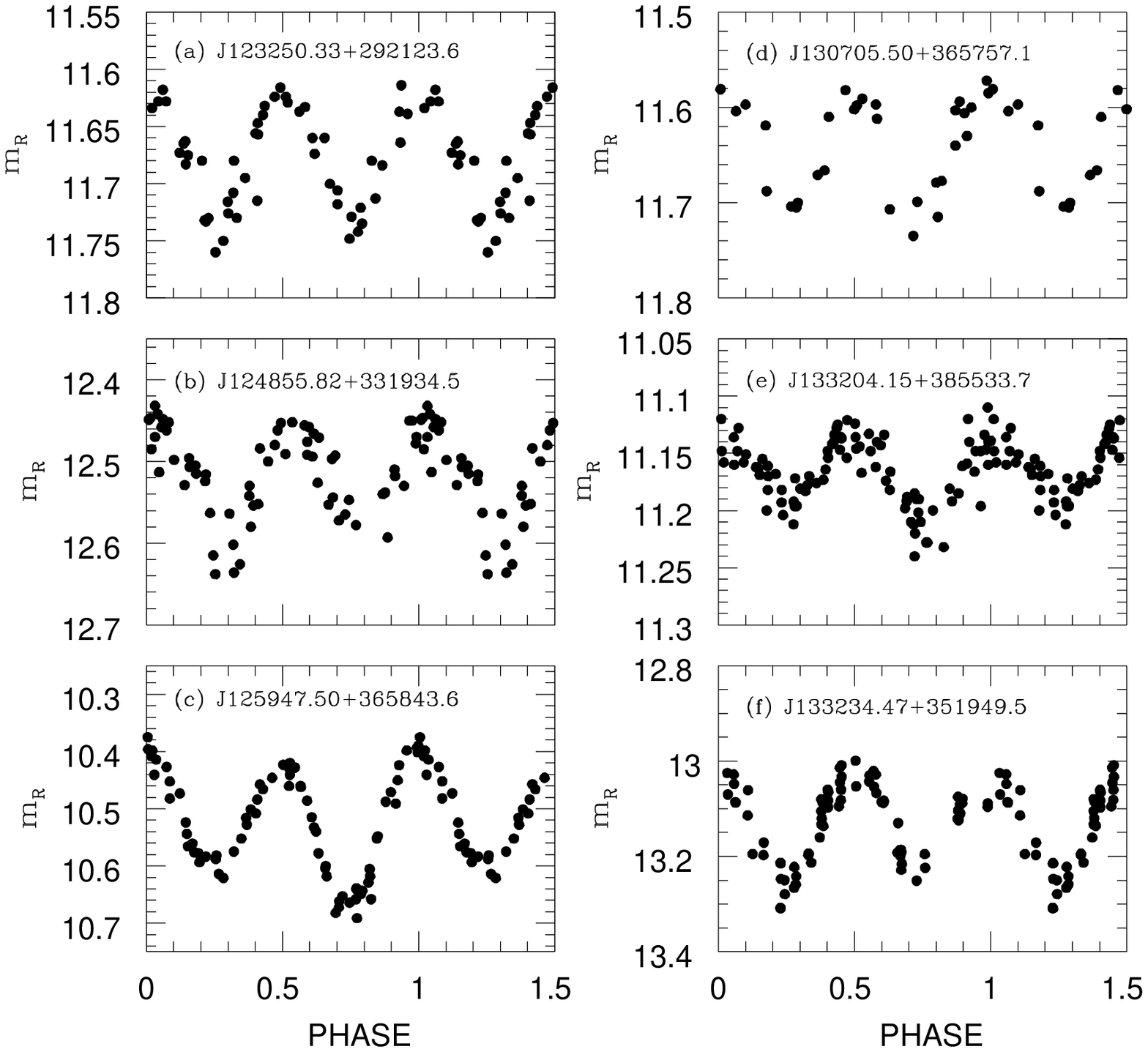}
\caption{ Light curves for J123250.33+292123.6, J124855.82+331934.5, 
         J125947.50+365843.6, J130705.50+365757.1,
         J133204.15+385533.7 and J133234.47+351949.5. 
         The $m_{R}$ magnitudes were taken  from the $NSVS$. 
         All these stars are reclassified as  eclipsing binaries.
\label{Fig5}}
\end{figure}

\subsection{Comments on stars classified as RR Lyrae stars.}  

 The classifications and periods derived from our photometry agree with
 those of AAB for the five stars with the highest amplitudes in Table 2.
 Their light curves are given in Fig. 3 and Fig. 4. 
 Some of the scatter in these  light curves may come from using 
 a single period for data that covers five years. Jurcsik et al. (2009) have 
 shown that light-curve modulation is a common property of RR0 (type {\it ab})
 stars and this may also contribute to the scatter (e.g. J123811.00+385028.0 
 where the modulation is very strong). Comments on the individual stars are
 given below:

 {\bf J123811.00+385028.0}: 
 The photoelectric data show that $(B-V)$ varies from 0.25 to 0.35 as would be 
 expected for an RR0 star. Our photoelectric and Tenagra data (2004, 2008 \&
 2009), however, were inadequate to give a satisfactory light curve. 
 In early 2009, we asked 
 Edward Schmidt to observe the star and he kindly made his observations
 available to us. The star clearly shows strong Blazhko effect (Fig. 3) but
 we have been unable to determine the Blazhko period. The adopted ephemeris
 is JD(max) = 2452407.588 and P = 0.533035 days. 

 {\bf J123854.21+245307.9}: Type RR0 (type {\it ab}). The adopted ephemeris is
  JD(max) = 2451244.905 and P = 0.628389 days with a 
   $V$ amplitude of 0.33 mag. and $(B-V)$ amplitude of 0.12.

 {\bf J130441.18+381805.1}: Type RR0 (type {\it ab}). The adopted ephemeris is
  JD(max) = 2451246.880 and P = 0.686270 days with a
   $V$ amplitude of 0.85 mag. and $(B-V)$ amplitude of 0.18.

 {\bf J132849.66+360757.1}: Type RR0 (type {\it ab}). The adopted ephemeris is  
  JD(max) = 2452407.365 and P = 0.586186 days with a
  $V$ amplitude of 0.73 mag. and $(B-V)$ amplitude of 0.10.

 {\bf J133048.36+335353.8}: This is the only confirmed RR1 (type {\it c}).
  The adopted ephemeris is
  JD(max) = 2451274.511 and P = 0.353000 days with a 
  $V$ amplitude of 0.42 mag. and $(B-V)$ amplitude of 0.15. 

 In all cases the light curves are significantly asymmetric and there is
 a significant color amplitude so that the RR classification seems secure.

\begin{figure}
\plotone{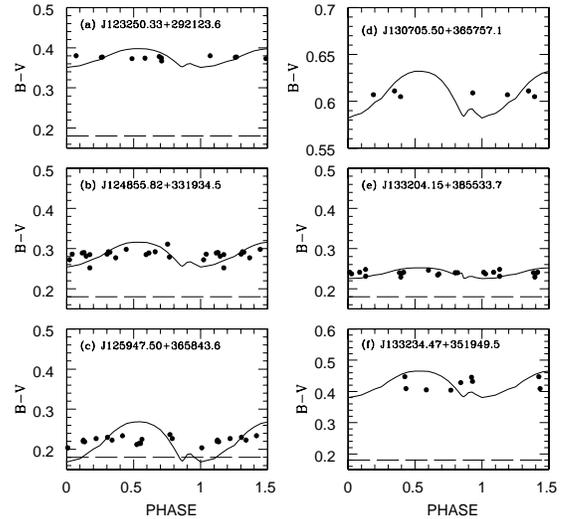}
\caption{ $(B-V)$ curves for J123250.33+292123.6, J124855.82+331934.5, 
         J125947.50+365843.6, J130705.50+365757.1,
         J133204.15+385533.7 and J133234.47+351949.5. 
         The $(B-V)$  colors are the Lowell (2002) observations.
         The curves are those expected for RRc variables and were obtained 
         by scaling the U Com colors (as in Fig. 2). The dashed horizontal
         line shows the mean $(B-V)$ of U Com which is generally much 
         bluer than those of these variables. 
         All these stars are reclassified as  eclipsing binaries.
\label{Fig6}}
\end{figure}

\subsection{Comments on stars classified as eclipsing binaries.}

 The remaining six stars in Table 2 
 have $m_{R}$ amplitudes of $\leq$ 0.3 magnitudes and were
 classified by AAB as RR1 (type {\it c}) variables. They do, however, have a
 very low amplitude or constant $(B-V)$ and so it is unlikely that they
 are pulsating stars. Support for this comes from their $m_{R}$ light-curves which
 use $NSVS$ magnitudes\footnote{As noted in Sec. 3, we omitted the $NSVS$ magnitudes
 that have large errors and combined those that have closely similar epochs.}.
 We used a periodogram program (Horne \& Baliunas  1986) to find the periods;
 these were close to those given by AAB.  We then doubled these periods before
 deriving the light curves (Fig. 5). In all cases, this double period gave two 
 minima of slightly unequal depth at phases 0.25 and 0.75 and the the maxima 
 were sometimes somewhat flattened.
 Fig. 6 shows the Lowell $B-V$ observations for these stars together with
 the color curves that would be expected if these stars were RRc. As in Fig.2,
 these curves were obtained by scaling the colors of U Com and the dashed
 horizontal line shows the mean $(B-V)$ color of U Com. The colors of these
 variables are generally significantly redder than that of U Com and do not
 show the phase variation of this RRc.

  Details for the individual variables are 
 given below: 

 {\bf J123250.33+292123.6}: (A) The adopted ephemeris is
  JD(max) = 2451244.603 and P = 0.524280 days. The $m_{R}$-amplitudes are
  roughly 0.110 and 0.125 magnitudes. 
  $<(B-V)>$ = 0.375 mag. and the $rms$ scatter in $(B-V)$ of a single 
  observation = 0.004mag.  

 {\bf J124855.82+331934.5}: (E) The adopted ephemeris is 
  JD(max) = 2451244.827 and P = 0.408280 days. The $m_{R}$-amplitudes are
  roughly 0.11 and 0.16 magnitudes. 
  $<(B-V)>$ = 0.286 mag. and the $rms$ scatter in $(B-V)$ of a single 
  observation = 0.012mag.  
  
 {\bf J125947.50+365843.6}: (G) The adopted ephemeris is
  JD(max) = 2451244.609 and P = 0.616285 days. The $m_{R}$-amplitudes are
  roughly 0.20 and 0.26 magnitudes. 
  $<(B-V)>$ = 0.222 mag. and the $rms$ scatter in $(B-V)$ of a single 
  observation = 0.009mag.  

 {\bf J130705.50+365757.1}: (I)  
  JD(max) = 2451244.614 and P = 0.44721  days. The $m_{R}$-amplitudes are
  roughly 0.11 and 0.13 magnitudes. 
  $<(B-V)>$ = 0.608 mag. and the $rms$ scatter in $(B-V)$ of a single 
  observation = 0.003mag. The $<B-V)>$ color is too red for this to be an
  RR Lyrae star.

 {\bf J133204.15+385533.7}: (M)  
  JD(max) = 2451244.790 and P = 0.578340 days. The $m_{R}$-amplitudes are
  roughly 0.05 and 0.07 magnitudes. 
  $<(B-V)>$ = 0.239 mag. and the $rms$ scatter in $(B-V)$ of a single 
  observation = 0.005mag.  

 {\bf J133234.47+351949.5}: (N)  
  JD(max) = 2451244.454 and P = 0.720306 days. The $m_{R}$-amplitudes are
  roughly 0.18 and 0.22 magnitudes. 
  $<(B-V)>$ = 0.424 mag. and the $rms$ scatter in $(B-V)$ of a single 
  observation = 0.018mag.  

 All the above stars are {\it reclassified}  as eclipsing binaries.

\begin{deluxetable*}{ cccccccccc }
\tablewidth{0cm}
\tabletypesize{\footnotesize}
\tablecaption{ Spectroscopic data for variables of Table 2.}                    
 
\tablehead{ 
\colhead{  $ROTSE1$ ID  } &
\colhead{  Type \tablenotemark{a}   } &
\colhead{  JD(hel.)\tablenotemark{b} } &
\colhead{ Int.   \tablenotemark{c}                } &
\colhead{ $\phi$ \tablenotemark{d} } &
\colhead{ R.V.           \tablenotemark{e}  } &
\colhead{ EW(H)$^{\ast}$    \tablenotemark{f}          } &
\colhead{ EW(K)$^{\ast}$    \tablenotemark{g}  } &
\colhead{$\Delta S$$^{\ast}$\tablenotemark{h}  } &
\colhead{ $\lambda$4481~W       \tablenotemark{i}  }  \\
                &    &        &    & &km$^{-1}$ &\AA\ &\AA\  &   &   \AA\    \\
}

\startdata
                   &    &                   &          &   &     &   &  &   &       \\
 RR Lyrae stars:   &    &                   &          &   &     &   &  &   &       \\
  J123811.00+385028.0&RR0&822.0584& 90& 0.665&--139& 07.5& 03.5& +3.5 & 1.8   \\
  J123854.21+245307.9&RR0&822.0550& 45& 0.573&--118& 06.5& 05.2& +2.3 & 1.9   \\
  J130441.18+381805.1&RR0&823.0601& 90& 0.039&--019& 09.2& 01.0& +7.5 & 1.2   \\
  J132849.66+360757.1&RR0&823.0568&120& 0.033&--272& 09.5& 02.3& +3.8 & 1.7   \\
  J133048.36+335353.8&RR1&823.0545& 90& 0.533&--128& 10.0& 01.4& +5.9 & 1.9   \\
 Eclipsing systems:&    &                   &          &   &     &   &  &   &       \\
  J123250.33+292123.6&EW  &823.0637& 60& 0.476& +-32& 10.1& 05.2&--3.8 &3.2    \\
  J124855.82+331934.5&EW  &822.0607& 75& 0.716&--065& 12.8& 03.6&--3.9 &3.2    \\
  J125947.50+365843.6&EW  &822.0635& 40& 0.064&--040& 14.4& 02.9&--5.4 &3.5    \\
  J133204.15+385533.7&EW  &920.0232& 60& 0.797& +004& 10.0& 03.7&--0.2 &2.6    \\
  J133234.47+351949.5&EW  &920.0210& 90& 0.564&--039& 03.9& 06.2& +3.7 &3.1    \\
                   &  &       &     &   & &      &    &              & \\
\enddata

\tablenotetext{a}{Type derived by our photometry: RRab = RR0; RRc = RR1     
 .} 
\tablenotetext{b}{Heliocentric Julian Date of mid-exposure (+2454000.0). } 
\tablenotetext{c}{Integration time in seconds.  } 
\tablenotetext{d}{Phase of mid-exposure. } 
\tablenotetext{e}{Heliocentric radial velocity in km s$^{-1}$. In the case of the RR Lyrae stars this has been corrected to the $\gamma$-velocity following            Liu (1991). } 
\tablenotetext{f}{Mean pseudo equivalent width of H$_{\gamma}$ \& H$_{\delta}$ (\AA).}
\tablenotetext{g}{Pseudo equivalent width of Ca {\sc ii} K-line (\AA).}
\tablenotetext{h}{Non-standard $\Delta S$ index.                   } 
\tablenotetext{i}{FWHM of Mg {\sc ii} $\lambda$~4481 line (\AA).     }

\end{deluxetable*}

\begin{deluxetable*}{ ccccccccccc }
\tablewidth{0cm}
\tabletypesize{\footnotesize}
\tablecaption{ Derived photometric and spectroscopic properties for stars in  
   Table 2.}
 
\tablehead{ 
\colhead{  $ROTSE1$ ID  } &
\colhead{  type \tablenotemark{a} } &
\colhead{ $\phi_{31}$ \tablenotemark{b}                } &
\colhead{[m/H]$^{\ast}$  \tablenotemark{c} } &
\colhead{[m/H]$^{\ast}$  \tablenotemark{d}  } &
\colhead{[m/H]$^{\ast}$  \tablenotemark{e}          } &
\colhead{[m/H]$^{\ast}$  \tablenotemark{f}  } &
\colhead{[m/H]$^{\ast}$  \tablenotemark{g}  } &
\colhead{$M_{v}$     \tablenotemark{h}  } &
\colhead{Dist.       \tablenotemark{j}  } &
\colhead{V           \tablenotemark{i}  } \\
                     &   &        &        &        &        &        &        &       &(kpc)    &km$^{-1}$    \\
}

\startdata
                     &   &        &        &        &        &        &        &       &         &             \\
 RR Lyrae stars:     &   &        &        &        &        &        &        &       &         &             \\
  J123811.00+385028.0&RR0&  2.39  &--0.96  &--0.74  &--0.84  &--0.66  &--0.8   &+0.69  &   3.70  & --180$\pm$055      \\
  J123854.21+245307.9&RR0&  2.67  &--0.77  &--0.85  &--1.01  &--0.80  &--0.8   &+0.68  &   2.89  & --189$\pm$047      \\
  J130441.18+381805.1&RR0&  1.94  &--1.6:  &--2.14  &--2.07  &--1.59  &--1.9   &+0.44  &   4.17  & --340$\pm$067       \\
  J132849.66+360757.1&RR0&  2.54  &--1.0:  &--0.82  &--1.35  &--0.90  &--1.0   &+0.65  &   6.02  & --435$\pm$300       \\
  J133048.36+335353.8&RR1&  2.43  &--1.36  &--1.8:  &$\cdots$&$\cdots$&--1.4:  &+0.57  &   2.75  & --115$\pm$044       \\
 Eclipsing systems:&     &        &        &        &        &        &       &        &         &              \\
  J123250.33+292123.6&EW &$\cdots$&$\cdots$&$\cdots$&$\cdots$&$\cdots$&$\cdots$& +2.43 & 0.62    &   +025$\pm$04         \\
  J124855.82+331934.5&EW &$\cdots$&$\cdots$&$\cdots$&$\cdots$&$\cdots$&$\cdots$& +2.68 & 0.79    &  --004$\pm$16         \\
  J125947.50+365843.6&EW &$\cdots$&$\cdots$&$\cdots$&$\cdots$&$\cdots$&$\cdots$& +1.68 & 0.49    &  --064$\pm$03         \\
  J130705.50+365757.1&EW &$\cdots$&$\cdots$&$\cdots$&$\cdots$&$\cdots$&$\cdots$& +3.46 & 0.37    & (--022$\pm$26)       \\
  J133204.15+385533.7&EW &$\cdots$&$\cdots$&$\cdots$&$\cdots$&$\cdots$&$\cdots$& +1.88 & 0.58    & --007$\pm$04          \\
  J133234.47+351949.5&EW &$\cdots$&$\cdots$&$\cdots$&$\cdots$&$\cdots$&$\cdots$& +1.99 & 1.45    & --072$\pm$16          \\
                     &   &        &        &        &        &        &        &       &         &             \\
\enddata

\tablenotetext{a}{Type derived by our photometry: RRab = RR0; RRc = RR1     
 .} 
\tablenotetext{b}{Fourier coefficient derived from hand-drawn light curve.  } 
\tablenotetext{c}{non-standard [m/H] derived from the spectra. } 
\tablenotetext{d}{non-standard [m/H] derived from the Fourier coefficient $\phi_{31}$. }
\tablenotetext{e}{non-standard [m/H] derived from the $V$ amplitude.} 
\tablenotetext{f}{non-standard [m/H] derived from the Rise-time (phase difference between minimum and following maximum). } 
\tablenotetext{g}{Adopted non-standard [m/H]$^{\ast}$.                             } 
\tablenotetext{h}{Absolute magnitude derived as described in text.}
\tablenotetext{i}{Distance in kpc. }
\tablenotetext{j}{Galactic velocity vector (V) in km s$^{-1}$ }

\end{deluxetable*}

\begin{deluxetable}{ ccccccccccc }
\tablewidth{0cm}
\tabletypesize{\footnotesize}
\tablecaption{ Distances and Galactic Rotations assuming the original
 classifications for the stars in Table 2.}
 
\tablehead{ 
\colhead{  $ROTSE1$ ID  } &
\colhead{  type \tablenotemark{a} } &
\colhead{$M_{v}$     \tablenotemark{b}  } &
\colhead{Dist.       \tablenotemark{c}  } &
\colhead{V           \tablenotemark{d}  } \\
                     &    &          &(kpc)    &km$^{-1}$    \\
}

\startdata
                     &   &           &         &                    \\
  J123811.00+385028.0&RR0&  +0.69    &   3.70  & --180$\pm$055      \\
  J123854.21+245307.9&RR0&  +0.68    &   2.89  & --189$\pm$047      \\
  J130441.18+381805.1&RR0&  +0.44    &   4.17  & --340$\pm$067       \\
  J132849.66+360757.1&RR0&  +0.65    &   6.02  & --435$\pm$300       \\
  J133048.36+335353.8&RR1&  +0.57    &   2.75  & --115$\pm$044       \\
                     &   &           &         &                      \\
  J123250.33+292123.6&RR1& +0.86 & 1.26    &   +053$\pm$09         \\
  J124855.82+331934.5&RR1& +0.86 & 1.90    &  --003$\pm$40         \\
  J125947.50+365843.6&RR1& +0.86 & 0.73    &  --093$\pm$07         \\
  J130705.50+365757.1&RR1& +0.86 &  1.28   & (--079$\pm$27)       \\
  J133204.15+385533.7&RR1& +0.86 & 0.98    & --012$\pm$06          \\
  J133234.47+351949.5&RR1& +0.65 & 2.85    & --135$\pm$33          \\
                     &   &           &         &                      \\
\enddata

\tablenotetext{a}{Type derived by AAB photometry: RRab = RR0; RRc = RR1     
 .} 
\tablenotetext{b}{Absolute magnitude derived from equation (1).}
\tablenotetext{c}{Distance in kpc. }
\tablenotetext{d}{Galactic velocity vector (V) in km s$^{-1}$ }

\end{deluxetable}

\begin{figure}
\plotone{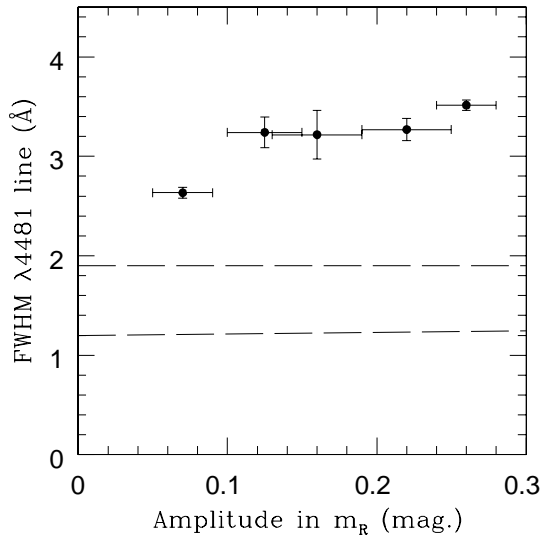}
\caption{ The ordinate gives the FWHM line-width in \AA~  of the 
      Mg {\sc ii} $\lambda$4481 doublet for the five 
      eclipsing systems for which spectra are available. The abscissa is
      the $m_{R}$-amplitude. For eclipsing binaries, this amplitude 
      should be least for 
      the stars that are seen nearly pole-on and which therefore should
      show the least line broadening.  The dashed horizontal 
      lines show the measured range of line widths for the RR Lyrae stars 
      for which the broadening should be very small.
\label{Fig7}}
\end{figure}

\section{Spectroscopy}  
   
  Spectra can be used to distinguish high-metallicity foreground eclipsing
  stars of the disk from low-metallicity variables that belong to the 
  extended low-rotation halo. 
  We obtained short integrations with the MMT blue-channel spectrograph of 
  all the stars in Table 2 except J130705.50+365757.1. Details of these
  flux-calibrated  spectra (that cover $\lambda\lambda$ 3600 -- 4500 \AA)
  are given in Table 5. The spectral resolution was 1.0 \AA\ for all the spectra
  except those of J133204.15+385533.7 and J133234.47+351949.5 for which it was
  1.2 \AA. The spectra have S/N in the range 50 to 100 and yield 
  radial velocities with accuracies of 2 to 3 km s$^{-1}$; the
  Balmer lines and Ca {\sc ii} K-line equivalent widths can be used to 
  determine the metallicity ([Fe/H). No calibrating standard stars were
  observed, however, so we could only measure pseudo equivalent widths from 
  which non-standard metallicities were derived. We denote these by 
  [m/H]$^{\ast}$. 

  \subsection{Metallicities and Rotations} 

  Preston (1959) showed that the difference between the spectral types of the
  Balmer lines and the Ca {\sc ii} K-line of RR Lyrae stars at minimum light 
  gives an index ($\Delta S$) that measures the metallicity of the star.
  These spectral types should be derived from the equivalent widths using a 
  calibration given by  spectra of stars of known spectral type that are taken
  concurrently with those of the program stars. Since we had no standards, we 
  used the equivalent width {\it vs.} spectral type calibration given by
  Kinman \& Carretta (1992); the resulting non-standard indices 
     (denoted by $\Delta S$$^{\ast}$)  were converted 
  to [m/H]$^{\ast}$  using the calibration of Suntzeff et al. (1994). 
  These metallicities are 
  approximate not only because of the rough calibration of the equivalent 
  widths but also because some of the spectra were taken at phases near
  maximum light where the conversion of $\Delta S$ to [Fe/H] is known to be
  inaccurate. Despite these limitations, the mean $\Delta S$$^{\ast}$ for the RR
  Lyraes (+4.6 $\pm$ 1.0) differs significantly from that of the eclipsing stars
  (--1.9 $\pm$ 1.8) in the sense that the RR Lyrae stars have halo 
  abundances ([Fe/H]$<$--0.8) while those of the eclipsing stars are more 
  nearly solar (Table 5).
  
   The metallicities of RR Lyrae stars can also be derived 
   from their light-curves.
   Smooth curves (drawn by eye through the data points) were used to derive 
  the Fourier coefficient $\phi_{31}$, the $V$-amplitude and the rise-time. 
   Non-standard metallicities [m/H]$^{\ast}$ 
   were  then derived from these quantities using the equations (3), (6) 
  and (7) respectively of Sandage (2004). The use of hand-drawn light curves
  necessarily gives approximate results but we see (Table 6) that these three 
  metallicity estimates  not only agree well among themselves but 
  also with the spectroscopic values. Our adopted  [m/H]$^{\ast}$ is based on 
  the three photometric and the spectroscopic values. Also, the [m/H]$^{\ast}$ 
   of --0.9 and --1.9 that we adopt for 
  J123854.21+245307.9 and J130441.18+381805.1 are in reasonable agreement 
  with the --1.2 and --2.2 respectively that Kinemuchi et al. (2006) 
  give for these stars. 

            We also measured the width 
  (FWHM of a gaussian fit) of the Mg {\sc ii} $\lambda$4481 doublet (Table 5). 
  This width is frequently used as a measure of rotational broadening 
  (Slettebak, 1954). In our data, this width is significantly larger for the
  eclipsing systems than for the RR Lyrae stars. The RR Lyrae stars are known   to have quite narrow lines (Peterson et al., 1996) while those of the 
  eclipsing systems should be broadened both by rotation and orbital motion.
  This  broadening should be least for those that are seen pole-on and thus
  for those 
  with the lowest amplitudes. It is seen in Fig. 7, that, in the case of
  the eclipsing stars, the FWHM of the Mg {\sc ii} $\lambda$4481 line does indeed
  decrease with the decreasing amplitude of the star. Extrapolated to zero
  amplitude, this width would be about 2.4 \AA. This may be compared with
  the mean width for the RR Lyrae stars
  (1.7$\pm$0.15\AA) which presumably is close to the instrumental width.
  
  \subsection{Absolute magnitudes, distances and Galactic rotation}.

  Absolute magnitudes for the RR Lyrae stars were obtained from:
  \begin{equation}  
    M_{v}  = 0.214[Fe/H] +0.86   
  \end{equation}  
  using the coefficients given by Clementini et al. (2003). Absolute 
  magnitudes for the eclipsing systems were obtained from the calibration 
  for W UMa-type binary stars given by Rucinski (2000): 
  \begin{equation}  
    M_{v}  = -4.44 \log P + 3.02(B-V)_{0} +0.12   
  \end{equation}  
  The distances given in Table 6 are based on these $M_{v}$ . Then,  
  following Johnson \& Soderblom (1987),  the radial velocities (Table 5)
  and proper motions (UCAC3 catalog: Zacharias et al., 2009) 
   were used to calculate the 
  heliocentric galactic rotation velocities (V) given in Table 6. 
  The radial velocities that we give for the RR Lyrae stars in Table 5 
  have been converted to $\gamma$-velocities following Liu (1991).
  As we noted above, the faintest of our RR Lyrae stars have UCAC3 proper
  motions whose errors are comparable with the proper
  motions themselves. Nevertheless, 
  the range in V that we find  for the RR Lyraes is quite different from 
  that which we find for the eclipsing stars; they correspond to what we 
  would expect for halo and disk stars respectively. 

  The galactic rotations in Table 6 are only valid if our reclassications
  are correct. We have therefore repeated the calculations  assuming that
  the six stars that we have reclassified as eclipsing binaries are in fact
  RRc stars as originally classified by AAB. The results are shown in Table 7.
  In this case, the mean galactic rotation of the six stars that we 
  reclassified is --45$\pm$31 km s$^{-1}$ and the dispersion in V for these
  stars is 69$\pm$20 km s$^{-1}$. {\it Thus, the galactic rotation (V) alone only 
  tells us that these six stars could be disk stars and either be eclipsing binaries
  or metal-rich disk RR Lyrae stars.}
  The most doubtful classification is
  that of J133234.47+351949.5 whose galactic rotation V and metallicity are
  compatible with it belonging to the halo. The large FWHM of its Mg {\sc ii}
  $\lambda$ 4481 line, however, makes it more likely that it is an
  eclipsing binary.   

  As would be expected for a disk population,
  the six stars that we reclassified as eclipsing systems have essentially
  zero mean radial velocity (--22$\pm$20 km s$^{-1}$). The RR Lyrae stars,
  on the other hand, show a mean radial velocity that is strongly negative 
  (--135$\pm$45 km s$^{-1}$). This downward streaming of halo stars at the
  North Galactic Pole was first discovered among subdwarfs by Majewski et al. 
  (1994, 1996). Similar  streaming in this part of the sky has since been also 
  found among RR Lyrae and and BHB stars (Kinman et al., 1996, 2006) and the
  velocities in Table 5 give further confirmation of the effect. This 
  downward motion towards the plane occurs a few kpc above the plane but
  does not continue to the solar neighborhood (Seagrove et al. 2008); it 
  presumably is one of the many tidal streams that are known to exist in 
  the halo.

\section{Summary}
 
 We use Johnson $BV$ photometry and MMT spectroscopy to study eleven stars
 (Table 2) classified by AAB as RR Lyraes. Our observations support the RR
 Lyrae classifications for the five stars with the largest amplitudes, however, 
 we find that the six low-amplitude stars classified by Akerlof et al. (2000)
 as RR1 (type $c$) should be reclassified as eclipsing 
 binaries. We derive metallicities from both spectra and light curves
 and find a halo abundance for the RR Lyrae stars and a solar abundance for
 the eclipsing binaries.

 The FWHM line widths of the Mg {\sc ii} $\lambda$4481 line provide a clean
 separation between RR Lyrae stars and eclipsing variables so that, in 
 principle, a single moderate S/N spectrum with a resolution of 1 \AA\ can
 provide enough information to distinguish between a pulsating variable and
 an eclipsing binary even when the amplitude is quite low. Supporting 
 evidence comes from the absence of color variation in eclipsing variables,
 and the differing kinematics of the two types of variables. 

 Our observations resolve the problematic over-abundance of RR1 (type $c$)
 reported by AAB. They found that 38\% of their RR Lyrae stars were RR1
 (type $c$), yet about half this number would have been expected from the Oo~I
  period distribution of these stars. In our 246 deg$^{2}$ field, there
 are 16 RR Lyrae stars with m$_{R}$ $\le$ 14.5. We find 4 RR1 (type $c$), 
 including BS Com discussed in Appendix A, and 12 RR0 (type $ab$). Thus in our
 small sample, 25\% of of the RR Lyrae stars are RR1 (type $c$); this is in
 statistical agreement with that expected from the period distribution.
 In the future, it would be desirable to investigate the classification of
 larger samples of the lower-amplitude RR Lyrae stars from the AAB test fields.
 Also many of the RR1 (type {\it c}) variables in the $ASAS$ 
 catalog are also classified as possible eclipsing stars; reclassification of 
 these stars using the techniques suggested here would also be very desirable.

\acknowledgments   

 We  thank the Director of the Lowell Observatory for allowing
 TDK to use the Lowell 42-inch telescope for this work and  also
 Dr David Schleicher (Lowell) both for help in 
 using the Kron photometer and in the preliminary reduction of the data. We 
 are also very grateful to Dr Przemek Wozniak (LANL) for his help with the 
 $ROTSE1$ catalogue. This research has made use of both the SIMBAD database
 and the VizieR catalogue access tool, operated at CDS Strasburg, France. 
 We also used the Two Micron All Sky Survey. which is a joint project of
 the Univ. of Massachusetts and IPAC (Cal. Tech.) and is funded by NASA and
 NSF. We also are grateful to Dr. Edward Schmidt for observing J123811.0+385028.0 
 and for allowing us to use his observations. We also thank the referee for comments 
 that have helped us to make significant improvements to the paper.

\clearpage

\appendix

\section{The survey field and previous surveys in this location}

\begin{deluxetable}{ cccccccc }
\tablewidth{0cm}
\tabletypesize{\footnotesize}
\setcounter{table}{8}
\tablecaption{ $ASAS$-3 variables with periods less than 1 day not identified
  by Akerlof et al., (2000). }  
 
\tablehead{ 
\colhead{  ID~($ASAS$-3)     } &
\colhead{ Class \tablenotemark{a} } &
\colhead{ Period         } &
\colhead{ $V_{max}$ \tablenotemark{b} } &
\colhead{ Amp             \tablenotemark{c}  } &
\colhead{ Other ID   \tablenotemark{d}  } &
\colhead{ Other Class \tablenotemark{e}  } &
\colhead{ $(J-K)_{0}$     \tablenotemark{f}  } \\
                   &         &(days)     &      &(mag.)  &    &       &        \\
}

\startdata
                   &         &     &      &        &    &       &        \\
 125534+2553.6 & MISC    & 0.3521 & 8.92 & 0.05 & IN~Com & R:/PN&0.508   \\
 131820+2452.3 & ED/DSCT & 0.4205 & 10.49 &0.50 &$\cdots$&$\cdots$ &0.329   \\
 131846+2547.5 & MISC    & 0.3304 & 9.86  &0.59 &$\cdots$&$\cdots$ &1.168   \\
 132656+2832.5 & MISC    & 0.6144 & 11.51 &0.23 &$\cdots$&$\cdots$ &0.657   \\ 
 133319+2300.7 & DSCT/EC & 0.1763 & 9.22  &0.09 &$\cdots$&$\cdots$ &0.378   \\
 133439+2416.6 & RRC:    & 0.3631 & 12.33 &0.80 & BS~Com & RRAB &0.231   \\
                   &         &     &      &        &    &       &        \\
\enddata

\tablenotetext{a}{Classification in $ASAS$-3: RRC = RR1, DSCT = Delta Scuti,    
 EC, ED = eclipsing system.} 
\tablenotetext{b}{$V$  magnitude at maximum light.} 
\tablenotetext{c}{$V$-amplitude in mag. } 
\tablenotetext{d}{Identification in $GCVS$ } 
\tablenotetext{e}{Variable type given in $GCVS$.  } 
\tablenotetext{f}{$2MASS$ color corrected for extinction.  }

\end{deluxetable}

 Fig. 1 shows the location of our survey field. The declination limits of
   +23.$^{\circ}$0 and +39.$^{\circ}$0 and the R.A. limit of 186.$^{\circ}$5
   were set by the limits of the test field of AAB (Akerlof et al. (2000)). The
   other R.A. limit (204.$^{\circ}$0) was chosen so that the field is roughly
   symmetrical about the NGP (R.A. 192.$^{\circ}$859, Dec. +27.$^{\circ}$128). 
    The RR Lyrae stars listed in the General Catalogue of Variable Stars 
   $(GCVS)$ are are shown as large open red pentagons.
    The remaining symbols (in black) show the stars that are listed by
    AAB according to the following AAB classifications:
   Filled circles  RR0 (type {\it ab}); open circles RR1 (type {\it c}); 
   open triangles Delta Scuti stars. The small filled squares are the stars
   that AAB classified as eclipsing binaries.

\begin{figure}
\setcounter{figure}{7}
\centerline{\includegraphics[width = 3.0in]{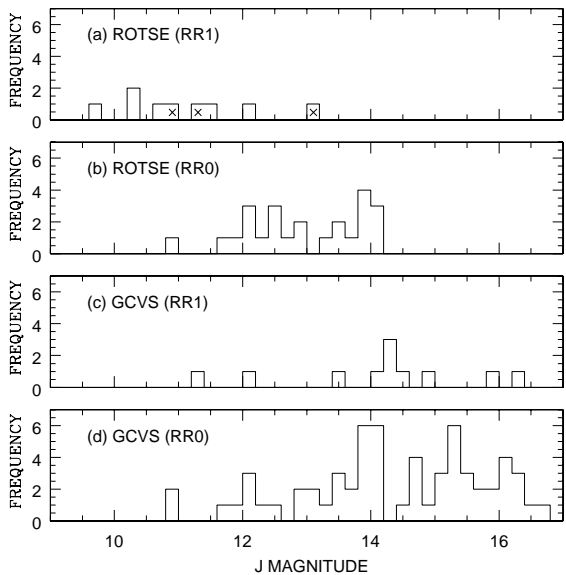}}
\caption{ The distributions  with $2MASS$ $J$ magnitude of RR0 (type $ab$) 
   and RR1 (type $c$) for the $ROTSE1$ survey (panels (a) \& (b)) and for 
   the $GCVS$ RR Lyrae stars (panels (c) \& (d)) in the field of
   Fig. 1. Unlike their $GCSV$ counterparts, the $ROTSE1$ RR1 cover a
    brighter range than the $ROTSE1$ RR0 stars. The three stars that we
    classify as RR1 (type {\it c}) in panel (a) are shown by crosses. We
    classify the rest of the stars in this panel as eclipsing binaries. 
\label{Fig8}}
\end{figure}

    Fig. 8 shows the distribution
   of RR0 (type {\it a}) stars and RR1 (type {\it c}) stars as a function of
   their $2MASS$ $J$ magnitude.  These distributions are shown separately for 
   the $ROTSE1$  and the $GCVS$ classifications. 
    { \it The  stars classified as RR1 (type {\it c}) are generally brighter
    than RR0 (type {\it ab}) in the $ROTSE1$ survey. This is not true for the
    $GCVS$ classifications.} In the top panel (a) of Fig. 8, we have markd
    with crosses the three stars that we classify as RR1 (type {\it c}); we
    classify the remainder as eclipsing systems. 

   The 81 deg$^{2}$ of our field South of declination +28.$^{\circ}$0 is also 
   covered by the $ASAS$-3 catalogue (Pojmanski  2002). Among the variables 
   with periods of less than a day in the two surveys, there are 
   six stars in $ASAS$-3 that are not given by AAB;
   they are listed in Table 8.  Two of these have amplitudes and colors that
   make them possible RR Lyrae stars. 131820+2452.3 is more likely to be a
   Delta Scuti star than an eclipsing system. The Northern Sky Variability
   Survey ($NSVS$) data between 1999 April and 1999 July show that the 0.21-day
   period gives a better light curve than the 0.42-day period (Fig. 10). With
   such a low amplitude ($\sim$0.1 mag.), 
   however,  this result is only provisional but the star is unlikely to be
   an RR Lyrae variable.
   133439+2416.6 (BS Com) has recently been shown to be a double-mode RR Lyrae
   star with fundamental and overtone periods of 0.487 and 0.363 days and
   amplitudes of 0.12 and 0.23 magnitudes respectively (Braggaglia et al. 
   (2003); Wils, (2006) and Szczygiel \& Fabrycky (2007)).
   D\'{e}k\'{a}ny (2007) has made detailed observations of BS Com; Fig. 4 
   in his paper shows how the amplitude of this variable varies through its
   cycle; an extensive set of data is needed to
   recognize such variables and this, perhaps, is why it was missed by AAB.

  Surveys for fainter RR Lyrae stars (roughly 12 $< V <$ 17) include the
  Sonneberg surveys which cover the same R.A. range as our field for declinations 
  South of +25.$^{\circ}$0 (Meinunger  1977) and the Survey with the Lick  
  Carnegie Astrograph (Kinman et al.  1966) in the region shown by the dotted 
  green lines in Fig. 1. Smaller deep surveys have been made by Pinto \& Romano
  (1973), Erastova (1979) and the CCD transit survey at +28.$^{\circ}$0  and
  width 8.2 arcmin by Wetterer et al. (1996). Most of the stars discovered in 
  these surveys are included in the $GCVS$ and  cover roughly 
   half of the total area of our field.
  The Lick Survey is known to be incomplete for variables with $B$ amplitudes
  less than 0.75 mag., and this is presumably true of these other
  {\it photographic} surveys. 
  In addition to the $ASAS$-3 survey, other new surveys for the brighter variables
   include the $SAVE$ CCD survey (Maciejewski \& Niedzielski  2005) 
   that covers the part of our region with R.A. $>$ 190.$^{\circ}$0
  and South of +29.$^{\circ}$0 and lists five red variables (9.0 $< V <$ 10.0)
  that are not listed in AAB.  The $SuperWASP$ survey 
  (Pollacco et al.  2006; Norton et al.  2007) started in 2004 and includes
  7.$^{\circ}$8 $\times$ 7.$^{\circ}$8 fields centered on Declination 
  +28.$^{\circ}$0 at all R.A.. It uses unfiltered CCDs and includes stars in the
  magnitude range 8 to 15; a catalog of the variables in this field is being
  compliled (Norton, {\it private communication}, May 2009) and should provide a  
  valuable supplement to the $NSVS$ catalog.

 The Northern Sky Variability Survey ($NSVS$) has also been searched for 
 RR Lyrae variables by Wils et al.  (2006) and by Kinemuchi et al.  (2006).
 In addition to the variables found by AAB,  Wils et al.
 re-discovered 
 the RR0 variables  WX CVn ($NSVS$ 7695165) and CY Com ($NSVS$ 7621236).

  There are no globular clusters in this field but M 3 is several tidal radii
  outside the field at R.A. = 205.$^{\circ}$546 and Dec = +28.$^{\circ}$.376. 
  The RR0 variables in this cluster have 15.49$ < V <$ 15.79 and the RR1 
  variables 15.27 $< V <$ 15.71 (Cacciari et al.  2005). These are too faint to be
  of concern in this paper.

\begin{figure}
\centerline{\includegraphics[width = 3.0in cm]{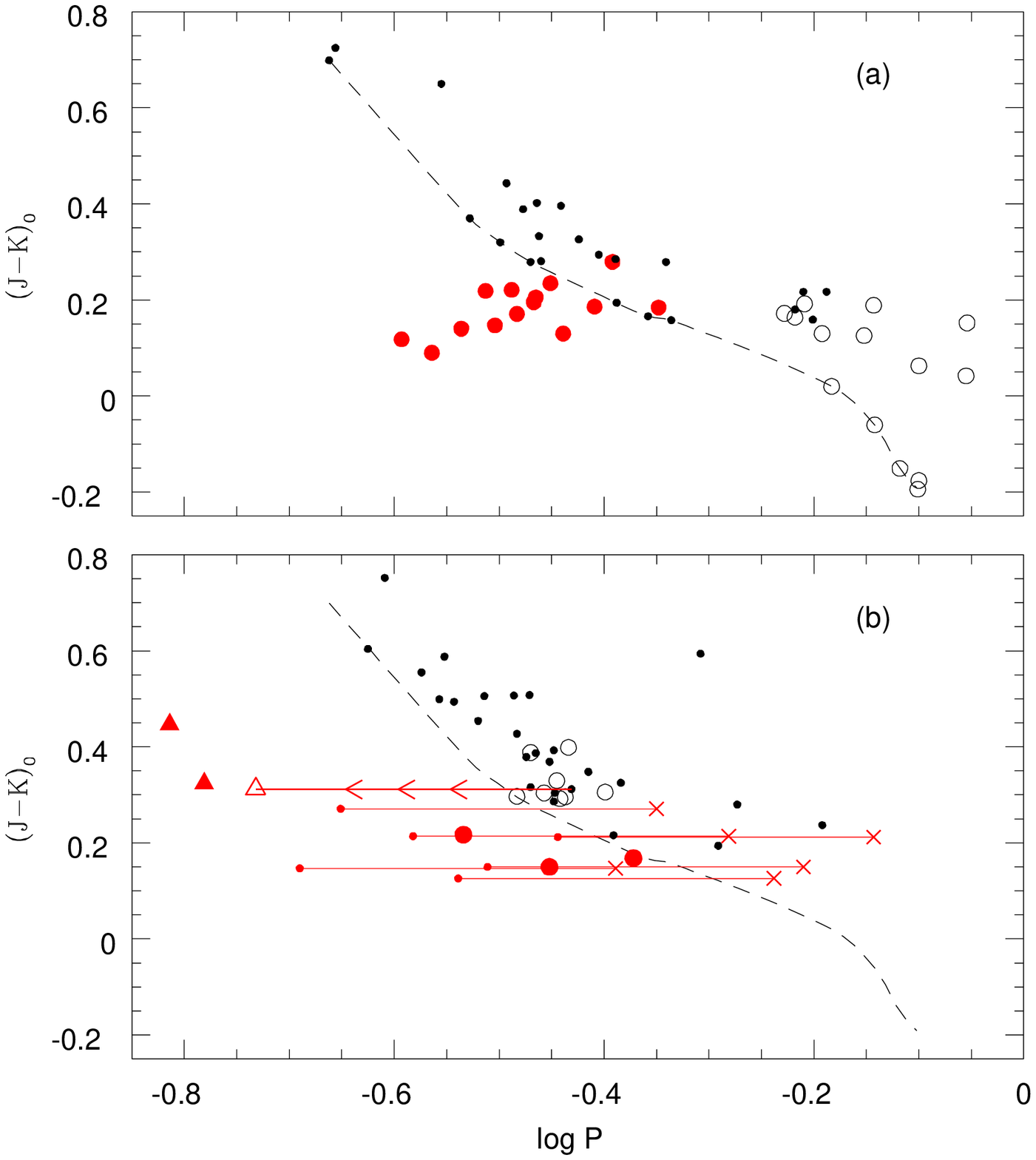}}
\caption{ ($(J-K)_{0}$ {\it vs.} logarithm of period (days) for (a) stars
   in the local neighborhood and (b) for stars in the field of Fig. 1. 
   Eclipsing systems are shown either by small filled or large
   open (black) circles. RR Lyrae stars shown by large filled red circles.
   Stars that we re-classified as eclipsing binaries are shown in (b) as
   red crosses. Delta Scuti stars are shown by triangles. Horizontal lines
   show the effect of doubling the period. The dashed curve is the
   approximate boundary between the eclipsing boundary and the RR Lyrae
   stars. 
\label{Fig9}}
\end{figure}

\begin{figure}
\centerline{\includegraphics[width = 3.0in]{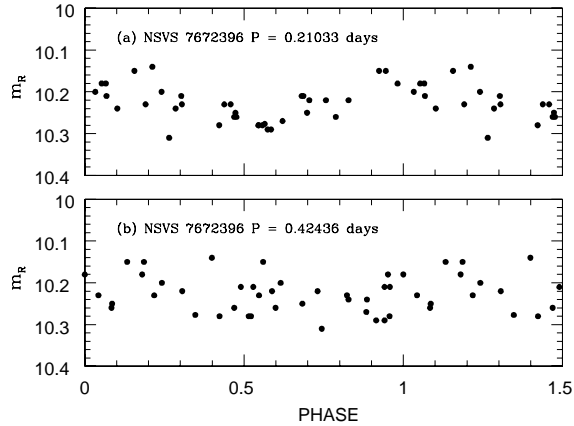}}
\caption{ Light curves for $NSVS$ 7672396 = $ASAS$-3 131820+2452.3.
 The $NSVS$ data (1999 April - July) show that a period of 0.21 days
   (shown above) is more likely than one of 0.42 days (shown below). 
\label{Fig10}}
\end{figure}

\section{On the confusion of RR1 (type {\it c}) variables with
         contact binaries in the color-period plot.}
 
 Contact binaries form a sequence in a color--period plot (Eggen  1961);
 Fig 9 shows this plot in the region where confusion between these variables
 and RR Lyrae and $\delta$-Scuti stars is most likely.
 Fig. 9(a) gives the plot of $(J-K)_0$ {\it vs.} log P in which nearby 
 contact binaries (Rucinski  2006; Rucinski \& Pribulla  2008) are shown as
 black filled circles and early type eclipsing systems (Eggen  1961; Eggen 
 1978) are shown as black open circles. In this paper, magnitudes are corrected
 for galactic extinction following Schlegel et al.  (1998).
  The black dashed curve shows the lower
 bound in this plot for the eclipsing systems. 
 Known nearby RR1 (type {\it c}) variables are shown in Fig 9(a)
 by red filled circles and there is a small overlap 
 between these variables and the eclipsing systems on this plot.
 Fig 9(b) shows the same plot for the $ROTSE1$ data. Here  
 two Delta Scuti stars are shown by red filled triangles. The red open
 triangle is a star ($V_{max}$= 11.80, $V-$amplitude 0.37 mag.)  at 
 13$^{h}$ 12$^{m}$ 29$^{s}$; +25$^{\circ}$ 14$^{'}$ 30$^{"}$ that is classified 
 as a Delta Scuti variable (Period 0.1854 days) in the $ASAS$-3  Survey
 (Pojmanski  2002) and as a eclipsing variable (P = 0.37089 days) in the $ROTSE1$
 Survey. The three stars that we have classified as RR1 (type {\it c}) are 
 shown by large filled red circles. The six stars that AAB classified as 
 RR1 would be located by the small filled circles if this classification were 
 correct and by the large crosses if they are (as we believe) eclipsing systems.
 It is seen that their reclassification as eclipsing systems relocates them in the
 color--period plot to a place that should be  well populated by eclipsing stars.

   
\section{The comparison stars}  

 Table 9 gives the positions and adopted magnitudes for the 
 comparison stars for the variables that are discussed in Sec. 2. 
 The positions are from the UCAC3 catalog (Zacharias et al., 2009)
 except for that of the comparison star of J133234.47+351949.5 which is
 from the $NOMAD$ catalog (Zacharias et al.  2004). The $(J-K)$ color
 is taken from the 2MASS catalog. 

\begin{deluxetable}{ ccccccc }
\tablewidth{0cm}
\tabletypesize{\footnotesize}
\setcounter{table}{9}
\tablecaption{ Coordinates and magnitudes of the comparison stars}

\tablehead{ 
\colhead{ Variable ID  } &
\colhead{ Comp. Star  } &
\colhead{  R.A. (2000)  \tablenotemark{a} } &
\colhead{ Dec. (2000)     \tablenotemark{b}  } &
\colhead{ $V$        \tablenotemark{c}  } &
\colhead{ $B$         \tablenotemark{d}  } &
\colhead{ $(J-K)$                        } \\
                   &    & (deg.)    & (deg.)    &      &     &               \\
}

\startdata
                   &    &           &           &      &     &               \\
 J123250.33+292123.6&A  & 188.1950  & +29.3739  &$\cdots$&14.11& 0.346     \\
 J123811.00+385028.0&A  & 189.5251  & +38.8471  & 13.84&14.47& 0.365         \\
                   & B  & 189.5493  & +38.8263  & 16.53&$\cdots$& 0.662      \\
 J123854.21+245307.9&A  & 189.7609  & +24.8800  & 12.88&13.59& 0.407         \\
 J124855.82+331934.5&B  & 192.1793  & +33.3364  &$\cdots$&15.97& 0.871       \\
                   & C  & 192.1750  & +33.3493  &$\cdots$&15.92& 0.349       \\
 J125947.50+365843.6&B  & 194.9810  & +36.9774  &$\cdots$&15.71& 0.546      \\
                   & C  & 195.0234  & +36.9740  &$\cdots$&15.72& 0.664       \\
 J130441.18+381805.1&A  & 196.1448  & +38.3304  & 13.85&14.52& 0.409         \\
                   & B  & 196.2226  & +38.3380  & 15.49&$\cdots$& 0.568      \\
 J130705.50+365757.1&A  & 196.8192  & +36.9761  &$\cdots$&14.13& 0.587       \\
                   & C  & 196.7330  & +36.9523  &$\cdots$&15.44& 0.234       \\
 J132849.66+360757.1&A  & 202.2527  & +36.0687  & 12.84&$\cdots$& 0.663      \\
                   & B  & 202.2017  & +36.1046  & 14.93&15.70& 0.539         \\
                   & C  & 202.2241  & +36.0924  & 14.97&15.56& 0.395         \\
                   & D  & 202.2494  & +36.0835  & 13.79&15.56& 0.615         \\
 J133048.38+335353.8&A  & 202.6450  & +33.8598  & 15.02&15.73& 0.351         \\
 J133204.15+385533.7&A  & 203.0287  & +38.9695  & 12.06&12.31& 0.109         \\
                   & B  & 203.0538  & +38.8945  & 14.48&15.02& 0.342         \\
 J133234.47+351949.5&A  & 203.1228  & +35.3107  & 16.48&17.54& 0.734         \\
\enddata

\tablenotetext{a}{R.A. in decimal degrees}     
\tablenotetext{b}{Dec. in decimal degrees}  
\tablenotetext{c}{$V$ magnitude } 
\tablenotetext{d}{$B$ magnitude } 

\end{deluxetable}

\end{document}